\documentclass[aps,prx,bibliography,twocolumn,showpacs,superscriptaddress,nofootinbib]{revtex4-1}
\usepackage{graphicx}
\usepackage{bm}
\usepackage{amssymb}
\usepackage{color}
\usepackage{amsmath}
\usepackage{amstext}
\usepackage{float}
\usepackage[outdir=./]{epstopdf}
\usepackage{latexsym}
\usepackage[usenames,dvipsnames]{xcolor}
\usepackage[colorlinks=true,citecolor=Blue,linkcolor=RubineRed,urlcolor=Blue]{hyperref}

\def\l{{\it  l}}

\def\la{\langle}
\def\ra{\rangle}
\def\om{\omega}

\def\tr{\rm{Tr}}

\def\dg{^\dagger}
\newcommand{\beq}{\begin{equation}}
\newcommand{\eeq}{\end{equation}}
\newcommand{\beqa}{\begin{eqnarray}}
\newcommand{\eeqa}{\end{eqnarray}}

\newcommand{\ket}[1] {\vert #1 \rangle}
\newcommand{\bra}[1] {\langle #1 |}
\newcommand{\braket}[2] {\langle #1 | #2 \rangle}
\newcommand{\Tr}{\mathrm{Tr}}

\newcommand{\aurelia}{\color{black}}

\graphicspath{{../images/}}

\begin{document}

\title{ Superadiabatic thermalization of a quantum oscillator by engineered dephasing}

\author{L. Dupays}
\affiliation{Donostia International Physics Center,  E-20018 San Sebasti\'an, Spain}
\affiliation{Institut d'Optique,  Palaiseau, Ile-de-France, France}
\author{I. L. Egusquiza}
\affiliation{Department of Theoretical Physics and History of Science,University of the Basque Country UPV/EHU, Apartado 644, E-48080 Bilbao, Spain}

\author{A. del Campo}
\affiliation{Donostia International Physics Center,  E-20018 San Sebasti\'an, Spain}
\affiliation{IKERBASQUE, Basque Foundation for Science, E-48013 Bilbao, Spain}
\affiliation{Department of Physics, University of Massachusetts, Boston, MA 02125, USA}
\author{A. Chenu}
\affiliation{Donostia International Physics Center,  E-20018 San Sebasti\'an, Spain}
\affiliation{IKERBASQUE, Basque Foundation for Science, E-48013 Bilbao, Spain}
\affiliation{Department of Chemistry, Massachusetts Institute of Technology, Cambridge, MA 02139, USA}

\begin{abstract}  Fast nonadiabatic control protocols  known as shortcuts to adiabaticity have found a plethora of applications,  but their use has been severely limited to speeding up the dynamics of isolated quantum systems. We introduce  shortcuts  for open quantum {\aurelia processes} that make possible the fast control  of Gaussian states in non-unitary processes. Specifically, we provide the time modulation of the trap frequency and dephasing strength that allow preparing an arbitrary thermal state in a finite time. Experimental implementation can be done via stochastic parametric driving {\aurelia or continuous measurements}, readily accessible in a variety of platforms.
 \end{abstract}
\maketitle

The fast control of quantum systems with high-fidelity is broadly acknowledged as a necessity to advance quantum science and technology. In this context, techniques known as shortcuts to adiabaticity (STA) have provided  an alternative to adiabatic driving with a wide variety of applications \cite{Torrontegui2013}. STA  tailor excitations in nonadiabatic processes to prepare a given state in a finite time, without the requirement of slow driving. The experimental demonstration of STA was pioneered in  a trapped thermal cloud \cite{Schaff2010}, soon followed by implementations in Bose-Einstein condensates \cite{Schaff2011a}, cold atoms in optical lattices \cite{bason2012} and low-dimensional quantum fluids \cite{Rohringer2015}. More recently, STA have also been applied to Fermi gases, both in the non-interacting and unitary regimes \cite{Deng18pra,Deng18Sci}.
Beyond the realm of cold atoms, STA  have  been demonstrated in quantum optical systems \cite{Du2016}, trapped ions \cite{An2016}, nitrogen vacancy-centers \cite{Zhang13}, and superconducting qubits \cite{Wang_2018,Zhang_2018}. Their application is not restricted to quantum systems and classical counterparts exist \cite{Jarzynski13,Deffner2014}, of relevance, e.g., to colloidal systems \cite{Schmiedl07}.

A variety of related control techniques fall under the umbrella of STA. Prominent examples include  counter-diabatic or transitionless quantum driving \cite{Demirplak2003,Demirplak2005,berry2009},  the fast-forward technique \cite{Masuda2009,Masuda2014}, reverse-engineered dynamics using Lewis-Riesenfeld invariants \cite{Chen10},  as well as the use of dynamical scaling laws \cite{Muga09,Campo2011,delcampo2013,Deffner2014}, Lax pairs \cite{Okuyama2016},  variational methods \cite{Sels2017} and  Floquet engineering  \cite{Claeys19}. The use of STA in the quantum domain is  severely limited to isolated  systems, in which sources of noise and decoherence are considered an unwanted perturbation \cite{Calzetta18,Levy_2018}. Applications to finite-time thermodynamics have thus been limited to the speedup of strokes in which the working substance is in isolation and decoupled from any external reservoir \cite{Deng18Sci}. 

Controlling heating and cooling processes would pave the way to the realization of superadiabatic heat engines and refrigerators based, e.g. in an Otto or Carnot quantum cycle \cite{Feldmann06,Deng13,Campo2014a,Beau16,Villazon19,Dann19}.
Hence, the possibility to speed up the dynamics of open quantum systems is highly desirable in view of applications to cooling, and more generally, in finite-time thermodynamics.
In this context,  the nonadiabatic control of  composite and open quantum systems using STA  remains an exciting open problem on which  few results are available \cite{Vacanti14,Duncan18,Dann18,Villazon19,Dann19,Alipour19}.

In this work, we introduce STA {\aurelia with open dynamics} and apply them to the superadiabatic cooling and heating of a thermal harmonic oscillator. We show that the required control protocols are local and involve only the driving of the trap frequency and the dephasing strength. They can be achieved using stochastic parametric driving, thus harnessing noise as a resource.

{\it Model.---} 
  We shall consider a single particle in a driven harmonic trap, with Hamiltonian
 \beqa \label{HO1}
\hat{H}_t=\frac{1}{2m}\hat{p}^2+\frac{1}{2}m\om_t^2\hat{x}^2,
 \eeqa
and {\aurelia a density matrix evolving according to} a  master equation of the form
 \beqa \label{eq:ME}
 \frac{d \rho_t}{dt}=-\frac{i}{\hbar}[\hat{H}_t,\rho_t]-\gamma_t[\hat{x},[\hat{x},\rho_t]],
 \eeqa
the derivation of which will be provided below.
 The case with constant dephasing strength $\gamma$  admits the Lindblad form with position operator $\hat{x}$ as the single  Hermitian Lindblad operator. This  naturally arises  as the high-temperature limit of quantum Brownian motion. 
 The dynamics with an arbitrary time-dependence $\gamma_t$ is generally non-Markovian. 
 We shall show how a time-inhomogeneous Markovian dynamics \cite{Rivas14}, corresponding to $\gamma_t>0$,  can be engineered by tailoring noise as a resource.
  The dynamics along the process is assumed to remain   Gaussian, with a density matrix in coordinate space of the form 
 \beqa
 \label{gaussansatz}
 \rho_t(x,x')=N_t \, e^{-A_t(x^2+x'^2)+iB_t(x^2-x'^2)-2C_txx'}, 
 \eeqa
 where $A_t,B_t,C_t$ are time-dependent coefficients to be determined from the master equation, $N_t= \sqrt{2 (A_t+C_t)/\pi}$ being the normalization factor.  This form includes coherences during the dynamics, and represents a family of dynamical processes that, as shown below, allows for a fast and controlled thermalization.
 
As a relevant example,   we consider the driving in a finite time $t_f$ of an initial thermal state, parameterized by the trap frequency and inverse temperature $(\om_0,\beta_0)$, to a different thermal state with $(\om_f,\beta_f)$.
For the Gaussian variational Ansatz (\ref{gaussansatz}) to describe the exact dynamics of the master equation (\ref{eq:ME}), the following consistency equations are to be satisfied (see App. \ref{appB}) 
\begin{subequations}
\label{consisteq}
\begin{align}
    \dot{B}_t&=\frac{2\hbar}{m}(A_t^2-B_t^2-C_t^2)-\frac{m\omega_t^2}{2\hbar},\\
  0&= \dot{A}_t+ \frac{4\hbar}{m}A_tB_t - \gamma_t =  \dot{C}_t+\frac{4\hbar}{m}B_t C_t+ \gamma_t.
\end{align}
\end{subequations}
The parameter   {\aurelia $2 \hbar B_t/m=-(\dot{A}_t+\dot{C}_t)/(2(A_t+C_t))\equiv \Omega_t$ is homogenous to a frequency, and directly follows from these equations.} 
The boundary conditions are given from the initial and final states, that we choose to be thermal. As detailed in App. \ref{appA}, it follows that  
$A_0=\frac{m\omega_0}{2\hbar}\coth(\hbar \om_0\beta_0)$, $B_0=0$ and $C_0=-m\omega_0/\big(2\hbar \sinh(\hbar\om_0\beta_0)\big)$.
Similarly, for the final state to be thermal, the coefficients at time $t_f$ should reduce to the values $A_f=\frac{m\omega_f}{2\hbar}{\rm coth}(\hbar \om_f\beta_f)$, $B_f=0$ and $C_f=-m\omega_f/\big(2\hbar \sinh(\hbar\om_f\beta_f)\big)$.
The initial and final states being taken as equilibrium states,  they are stationary. This imposes  the additional boundary conditions  $\dot{A}_{0/f}=\dot{B}_{0/f}=\dot{C}_{0/f}=0$. We also require that  
$\ddot{A}_{0/f}=\ddot{B}_{0/f}=\ddot{C}_{0/f}=0$ at initial and final time---the latter conditions are auxiliary, but  guarantee a smooth variation of $\om_t$ and $\gamma_t$.  

A protocol  speeding up the evolution from the thermal state characterized by 
$(\om_0,\beta_0)$ to $(\om_f,\beta_f)$ is obtained by   explicitly specifying both the time-dependence of $\gamma_t$ and $\om_t$, as directly given by the consistency equation (\ref{consisteq}), according to
 \begin{eqnarray}
   \omega^{2}_t&=&\frac{4\hbar^{2}}{m^{2}}(A^{2}_t-C^{2}_t)-\frac{3}{4}\left(\frac{\dot{A}_t+\dot{C}_t}{A_t+C_t}\right)^{2}+\frac{1}{2}\frac{\ddot{A}_t+\ddot{C}_t}{A_t+C_t}\label{eq:omega},\\
    \gamma_t&=&\frac{\dot{A}_tC_t-A_t \dot{C}_t}{A_t+C_t}.
\end{eqnarray}
Engineering a shortcut to thermalization between Gaussian states thus requires the ability to control both the frequency and dephasing. The control of the harmonic frequency $\omega_t$ is performed with routine in a variety of setups and has been used to implement STA in isolated quantum systems, e.g., with trapped ultracold atomic systems \cite{Schaff2010,Schaff2011a,Rohringer2015,Deng18pra,Deng18Sci}. The requirement of  a time-dependent dephasing $\gamma_t$ makes the dynamics open. It can be experimentally implemented from the microscopic picture provided below.

{\it Engineering of time-dependent dephasing rates.---}
To modulate the dephasing strength $\gamma_t>0$ in the laboratory we propose {\aurelia two different strategies}: (i) harnessing noise as a resource \cite{budini2001,Chenu2017b} {\aurelia or (ii) via continuous measurements, which have been implemented in e.g. trapped ions \cite{Smith18} and solid-state qubits  \cite{korotkov1999}, respectively.}
 
(i) {\aurelia The master equation (\ref{eq:ME}) can be obtained from implementing} the stochastic Hamiltonian 
\begin{equation} \label{eq:Htilde}
\hat{H}_{\rm st} = \hat{H}_t +  \hbar \sqrt{2 \gamma_t} \xi_t \hat{x} ,
\end{equation}
characterized by the Wiener process $W_t = W_0 + \int_0^t \xi_{t'} dt'$ defined in terms of the real Gaussian process $\xi_t$. While such a  stochastic process is not differentiable, all integral quantities can be defined from the Wiener increment $dW_t = \xi_t dt$. The noise-averaged expressions follow from the moments $\la \xi_t \ra$ and $\la \xi_t \xi_{t'} \ra$, that we choose  to be zero and $\delta(t-t')$, respectively,  to describe a real Gaussian white-noise process \cite{CarmichaelBook1}.  

The evolution of a quantum state dictated by the stochastic Hamiltonian (\ref{eq:Htilde}) is described by a master equation that we derive below. For a small increment of time $dt$, the wave function can be written as $
\ket{\psi_{t+dt}} = \exp\big(- i (\hat{H}_t dt/\hbar + \sqrt{2\gamma_t} \hat{x} dW_t)\big) \ket{\psi_t}$, with 
 $dW_t$  defined in the It\^o sense, i.e. fulfilling $(dW_t)^2 = dt$ and $dW_t dt=0$ \cite{Adler2003a, GardinerBook,Ruschhaupt2012a}.
A Taylor expansion of the exponential then gives  
\begin{equation}
d\ket{\psi_t} = \Big( -\frac{i}{\hbar} (\hat{H}_t dt +\hbar \sqrt{2\gamma_t} \hat{x} dW_t) - \gamma_t \hat{x}^2 dt \Big)\ket{\psi_t},
\end{equation}
  the only non-zero terms being first order in $dt$ or $(dW_t)^2$. 
Further, in the It\^o calculus, the Leibnitz chain rule generalizes to $d(AB) = (A+ dA) (B+dB) - AB = (dA) B + A (dB) + dA dB $. 
This gives the evolution of the density matrix $\rho_{\rm st} = \ket{\psi_t}\bra{\psi_t}$ as 
\begin{equation}
d \rho_{\rm st}= - \frac{i}{\hbar} [\hat{H}_t, \rho_{\rm st}] dt - i \sqrt{2\gamma_t} [\hat{x},\rho_{\rm st} ]dW_t-\gamma_t [\hat{x},[\hat{x},\rho_{\rm st} ]]dt,
\end{equation}
which preserves the norm at the level of each individual realization.
We then take the average over the realizations of the noise, and denote the ensemble $\rho_t= \la \rho_{\rm st}\ra$. Using the fact that the average of any function $F_t$ of the stochastic process vanishes, $\la F_t dW_t\ra=0$ \cite{Adler2003a}, we find that the evolution for the ensemble density matrix $\rho_t$ as dictated by the master equation (\ref{eq:ME}).  

{\aurelia 
(ii) Alternatively, the same evolution can be induced via continuous quantum measurements \cite{jacobs2006, korotkov1999, Bookwiseman2009, jacobs2014, jacobs2007} whenever the strength of the measurement is time-varying.    
Consider a quantum system subject to a continuous quantum measurement of the observable $\hat{A}$. Its evolution is known to be described by the stochastic non-linear master equation 
 \cite{Bookwiseman2009,jacobs2014}
\beqa
d\rho_t^{st}=L(\rho_t^{st})dt+I(\rho_t^{st})dW_t,
\eeqa
where $dW_t$ denotes a random Gaussian real random variable of zero mean and variance $dt$. The characteristic measurement time with which observable $\hat{A}$ is monitored, denoted $\tau_m$,  can be controlled by changing the measurement strength. The deterministic part of the evolution $L(\rho_t^{st})$ includes a non-unitary term of the standard Lindblad form,
\beqa
L(\rho_t^{st})=-\frac{i}{\hbar}[\hat{H}_t,\rho_t^{st}]-\frac{1}{8\tau_m}[\hat{A},[\hat{A},\rho_t^{st}]],
\eeqa
while the so-called innovation term reads
$I(\rho_t^{st})=\sqrt{\frac{1}{4\tau_m}}\big(\{\hat{A},\rho_t^{st}\}-2\Tr(\hat{A}\rho_t^{st})\rho_t^{st}\big)$. 
The latter is non-linear in the state $\rho_t^{st}$ and represents the measurement back-action on the system resulting from the acquisition of information during the measurement process.
A  specific trajectory is associated to   a given realization of the Wiener process $dW_t$, and characterized by fluctuations of the 
measurement outcomes given by $dr_A=\la \hat{A}\ra(t)+\sqrt{\tau_m}dW_t$.
When the observer does not have access to the measurement outcomes, the system is consistently described by the state $\rho_t=\la\rho_t^{st}\ra$, which results from averaging over an ensemble of trajectories, and that satisfies 
$d\rho_t=L(\rho_t)dt.$  
So  monitoring the position operator ($\hat{A}=\hat{x}$) with a time-dependent measurement strength such that $\frac{1}{8\tau_m}=\gamma_t$ (obtained e.g. by applying feedback on the system 
  \cite{jacobs2007}), effectively generates the master equation (\ref{eq:ME}). 
}

To sum up, the engineering of a prescribed modulation in time of the dephasing strength $\gamma_t$ can  be achieved via stochastic parametric driving or continuous measurements, provided that $\gamma_t > 0$. Interestingly, both techniques allow  modulating $\gamma_t$ independently from the frequency $\omega_t$, which contrasts with the time-dependent Markovian quantum master equation derived by  driving the coupling of a system to a thermal bath \cite{Dann18b}. 
{\aurelia Our scheme can be readily implemented in a single trapped ion \cite{leibfried2003}, in which the creation of an open dynamics with artificial environment  \cite{myatt2000, turchette2000}  or via the addition of noise \cite{Smith18} have been experimentally demonstrated.} 
  
  {\it Characterization of the dynamics.--- }
   The evolving density matrix can be diagonalized at all time according to $\rho_t=\sum_n p_{n,t} \ket{\psi_{n,t}}\bra{\psi_{n,t}}$,  
  the eigenvalues and eigenfunction being (see App. \ref{appC} and \cite{Mehler1866}) 
 \begin{subequations}
\begin{align}
\braket{x}{\psi_{n,t}} &= \sqrt{\frac{k_t}{2^n n! \sqrt{\pi}}} e^{-\frac{k_t^2}{2} x^2} e^{ i B_t x^2}{H}_n (k_t x)
\\
p_{n,t} &= u^n_t (1-u_t) ,
\end{align}
\end{subequations}
where ${H}_n$ denotes the Hermite polynomial defined from $\frac{d^j}{dx^j}({H}_N(x) ) = 2^j N!/(N-j)! {H}_{N-j}(x)$. The effective inverse length $k_t$ and  dimensionless  constant $u_t$ that characterize the control trap are detailed in App. \ref{appC} and below.
 Interestingly, the evolving density matrix $\rho_t$ can be interpreted as a thermal state $\sigma_t$ rotated through a unitary transformation  $\hat{U}_{x,t}\equiv e^{-i B_t \hat{x}^2}$  by noting that 
 \begin{equation} \label{rhosigma}
 \rho_t = \hat{U}_{x,t}^\dagger \sigma_t \hat{U}_{x,t}.
 \end{equation}
  The density matrix $\sigma_t$, with coordinate representation $\bra{x} \sigma_t \ket{x'} = N_t e^{- A_t (x^2 + x'^2) - 2C_t xx'}$, corresponds to the instantaneous thermal state of a harmonic oscillator with effective frequency $\tilde{\omega}_t$ and inverse temperature $\tilde{\beta}_t$ provided that  
 \begin{eqnarray} \label{eq:wtilde}
  \tilde{\varepsilon}_t \equiv \tilde{\beta}_t \hbar \tilde{\omega}_t &=&  {\rm acosh}(-A_t/C_t), \\
\tilde{\omega}_t^2 &=& \frac{4 \hbar^2}{m^2} (A_t^2-  C_t^2), 
\end{eqnarray}
assuming oscillators of equal mass. The effective inverse length is then   explicitly given by  $k_{t} = \sqrt{m \tilde{\omega}_t / \hbar}$. 
 By construction, the two states share the same eigenvalues and $\sigma_t = \sum_n p_{n,t} \ket{n_t}\bra{n_t}$, with the probability now written in terms of a thermal probability at all times, $p_{n,t}= e^{- \tilde{\beta}_t \hbar \tilde{\omega}_t n}/Z_t$, the partition function being $Z_t=1/(1-u_t)$, and  $u_t = e^{-\tilde{\varepsilon}_t}$. However, the eigenvectors are different and $\ket{n_t} = \hat{U}_{x,t} \ket{\psi_{n,t}}$ correspond to the well-known Fock states of the `reference', time-dependent harmonic oscillator $\tilde{H}_t$---whose parameters are distinguished with a tilde. 

 At all times of evolution, we have $A_t = (k_{t}^2/2) \coth \tilde{\varepsilon}_t$ and $C_t = -k_{t}^2/(2  \sinh\tilde{\varepsilon}_t)$, which lead to 
 {\aurelia  $\Omega_t = -\frac{1}{2} \frac{\dot{\tilde{\omega}}_t}{\tilde{\omega}_t} + \frac{\dot{u}_t}{1-u_t^2}$}. So the control frequency and dephasing strength  can be recast in the form   
 \begin{eqnarray} \label{control1}
 \omega_t^2 &=&  {\aurelia \tilde{\omega}_t^2 - \Omega_t^2 - \dot{\Omega}_t}=  \tilde{\omega}_t^2 - \frac{\ddot{\eta}_t}{\eta_t}, \label{eq:omega}\\
  \gamma_t &=&k_{t}^2 \frac{\dot u_t}{(1-u_t)^2}=-k_{t}^{2}\frac{ \dot{\tilde{\varepsilon}}_t}{4\sinh^2(\tilde{\varepsilon}_t)},\label{eq:gammat}
 \end{eqnarray}
 where the control parameter depends on the scaling factor $\kappa_t \equiv k_0 / k_t = \sqrt{\omega_0 / \tilde{\omega}_t} $ and temperatures as 
 \begin{equation}\label{eq:eta}
\eta_t=N_0 /N_t  =\kappa_t \sqrt{\coth(\tilde{\varepsilon}_0/2)\tanh(\tilde{\varepsilon}_t/2)}. 
 \end{equation}
 
 These are our main results. The combined modulation of the trap frequency and the dephasing strength is sufficient to engineer  finite-time shortcuts to thermalization. Equation (\ref{eq:omega}) gives the correction of the control of the  trap frequency  $\omega_t$ with respect to a reference one $\tilde{\omega}_t$ that needs to be experimentally implemented  for the preparation of the thermal state in a finite, prescribed time. A comparison of these results with the ones reported for isolated systems with $\gamma_t=0$ \cite{delcampo2013} shows that the control parameter in Eq.~(\ref{eq:eta})  not only depends on the scaling factor, but also accounts for the change of temperature through an additional, non trivial term. 
 
 \begin{figure}
\includegraphics{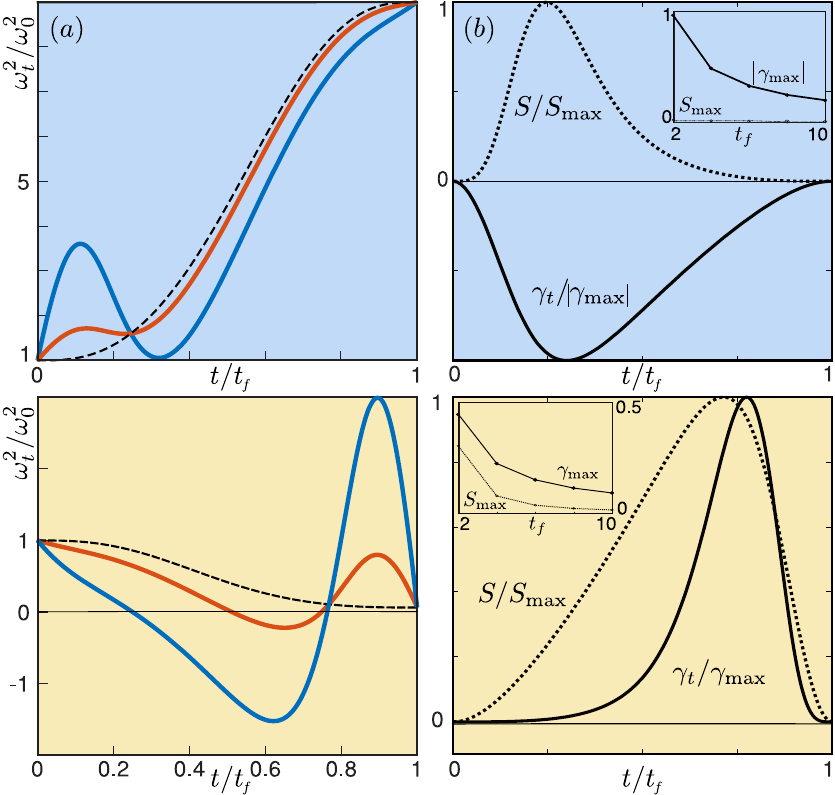}
\caption{(a) Control frequency  $\omega^2_t/\omega^2_0$  for protocols with duration  $t_f = 2$ (blue) and  6 (red), and reference frequency $\tilde{\omega}^2_t/\omega^2_0$ (dashed black). Fast protocols can require an inversion of the trap. (b) Dissipation rate  $\gamma_t/\gamma_{\rm max}$ (solid lines)  used to control the dynamics and  relative entropy $S(\rho_t\|\sigma_t)/S_{\rm max}$ (Eq. (\ref{eq:srel}), dotted lines), showing the distance of the state  to a reference thermal state.   The shape of the dephasing strength and the relative entropy is independent on the duration of the protocol, which only influences their maxima. Protocols correspond to cooling, $\beta_{f} / \beta_{0}=2$, with compression (top, $\omega_{f} /\omega_{0}=3$) or expansion (bottom, $\omega_{f} / \omega_{0}=1/4$), with $\omega_{0}=\beta_{0}=1$.
}\label{fig1}
\end{figure}

 Our scheme can be implemented by choosing an interpolating Ansatz between the boundary conditions imposed on the state, i.e. $\omega_{0/f}$ and $\beta_{0/f}$ that define $A_{0/f}$ and $C_{0/f}$. 
For illustration, we choose
 \begin{subequations}
\label{eq:polynome}\begin{align}
A_t &= A_0 + (A_f - A_0) \,f(t/t_f) {\rm  \, and } \\
C_t &= C_0 + (C_f -C_0) \, f(t/t_f) 
\end{align}
\end{subequations}
 in the form of a fifth-order polynomial, $f(\tau) = 10\tau^3 - 15 \tau^4+6 \tau^5$, to ensure a smooth dynamics.
The modulation of the control parameters  $\omega_t^2$ and $\gamma_t$ readily follow from Eqs.~(\ref{eq:wtilde}-\ref{eq:eta}). Fig.~\ref{fig1} illustrates the control frequency and dephasing strength corresponding to phase-space compression and expansion protocols, discussed below. Short control processes require trap inversion (a negative squared frequency), which can be achieved experimentally via, e.g.,   a painted potential \cite{Henderson09} or a digital micromirror device \cite{Gauthier16}. They also rely on a dephasing strength of larger amplitude, which can be experimentally more challenging to achieve. We propose to use the maximum of the dephasing strength, denoted $\gamma_{\rm{max}}$ as a measure of the ``cost'' to  implement the process by a technique such as stochastic parametric driving. 
We show in App. \ref{appD} that the maximum dephasing strength scales inversely with the process time, as illustrated in the insets of Fig. \ref{fig1}b. 

We further use  the relative entropy, defined as $S(\rho_t \|\sigma_t) = \Tr\big(\rho_t \ln \rho_t\big) - \Tr\big(\rho_t \ln \sigma_t\big)$, as a measure of the distance of the engineered state $\rho_t$ to the effective thermal state $\sigma_t$ along the dynamics. It can be written as  
  \begin{equation}\label{eq:srel}
S(\rho_t \|\sigma_t) = \sum_{n=0}^\infty p_n  \ln p_n  - \sum_{m=0}^\infty ( p_n \ln p_m) \big|\braket{m_t}{n_t}\big|^2 , 
 \end{equation} 
where the overlap of the eigenfunctions $\braket{m_t}{n_t}$ is given explicitly in App. \ref{appE} following \cite{delCampo2017a, Chenu2019b, AndrewsBook}. Fig.~\ref{fig1} illustrates the  relative entropy between the engineered and the thermal state. The insets show the maximum relative entropy for different process times, evidencing that the state is going further away from a thermal distribution for shorter protocols. 
The shape of the dephasing strength and the relative entropy is independent on the duration of the process, which only influences their maxima, $\gamma_{\rm max}$ and $s_{\rm max}$, respectively, as illustrated in Fig.~\ref{fig1}b.

 \begin{figure}
\includegraphics{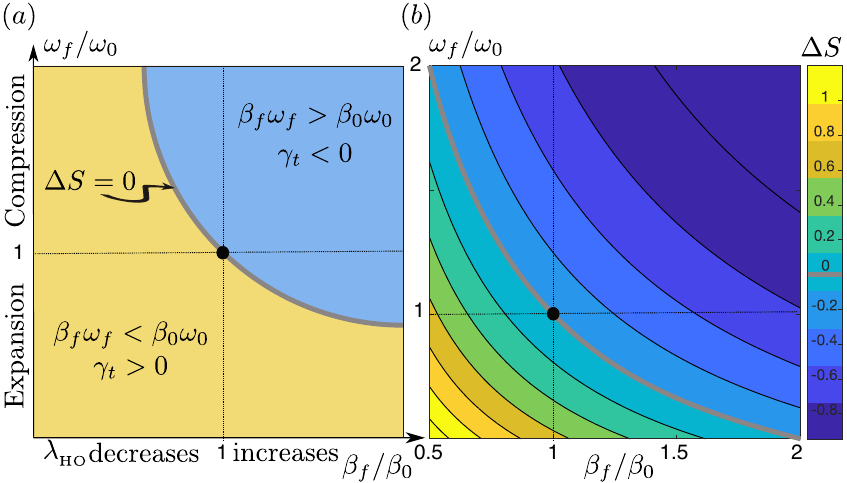}
\caption{(a) Schematic representation of the phase-space map of final thermal states ($\beta_f, \omega_{f}$) reachable from an initial thermal state ($\beta_0$, $\omega_0$), highlighted with the black dot, through STA control processes. States for which  $\beta_{f}\omega_{f}=\beta_0\omega_0$, chosen along the gray line, can be obtained from unitary, phase-space preserving processes, with no entropy change. STA in open processes allow preparing arbitrary thermal states, while stochastic driving with $\gamma_t>0$ gives access to states  for which $\beta_{f}\omega_{f}<\beta_{0}\omega_{0}$ (region highlighted in orange). (b) Associated change of the von Neumann entropy for $\omega_{0}=\beta_{0}=1$.} \label{fig2}
\end{figure}

{\it Superadiabatic protocols.---}  
Processes satisfying  $\beta_{f}\omega_{f}=\beta_{0}\omega_{0}$
 conserve the mean phonon number and are often referred to as 
 phase-space (density) preserving. The inverse temperature and frequency can be related to two physical lengths, namely, the particle characteristic length, given by the de Broglie wavelength, $\l_{\rm{dB}}=\hbar\sqrt{\beta/(2m)}$, and the trap characteristic length, $\lambda_{_{\rm  HO}}=\sqrt{\hbar/(m\omega)}$. Their ratio $l_{\rm{dB}} / {\lambda_{\rm HO}} = \sqrt{\beta \hbar \omega /2}$ is conserved for  phase-space preserving transformation. 
  STA in closed systems are limited to phase-space preserving cooling techniques, such as adiabatic cooling. These processes preserve the von Neumann entropy $S_t = -\tr(\rho_t\ln\rho_t)$ .  By contrast,  cooling and heating processes altering the phase-space density and the number of populated states lead to an entropy change \cite{ketterle1992} and  require an open dynamics. 
  \begin{figure} 
 \centering
\includegraphics{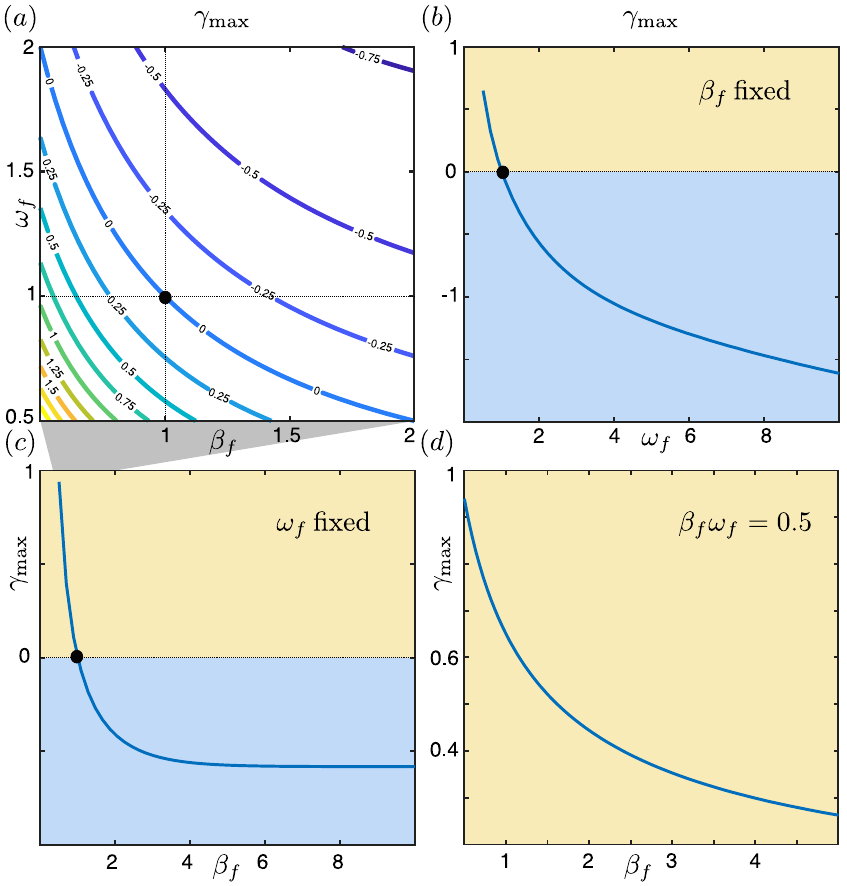}
\caption{ Maximum dephasing strength $\gamma_{\rm{max}}$: (a) Contour plot  for an initial state $\beta_{0}=1=\omega_{0}$. We verify that $\gamma_{\rm{max}}$ is negative for $\beta_{f}\omega_{f}>\beta_{0}\omega_{0}$,  and nonnegative otherwise; 
 (b)  as function of $\omega_{f}$ for fixed $\beta_{f}=1$; (c) as function of $\beta_{f}$, keeping the trap frequency unchanged ($\omega_f = 1$), showing a plateau for large values of $\beta_{f}$; (d) and on a line of constant $\beta_{f}\omega_{f}=0.5$ showing that $\gamma_{\rm{max}}$ depends on the specific choice of $\beta_{f}$ and $\omega_{f}$.}
\label{fig3}
\end{figure}

STA for open processes thus allow reaching arbitrary   thermal states ($\omega_{f}$, $\beta_{f})$  from an initial thermal state, as schematically represented in Fig.~\ref{fig2}, along with the variation of entropy. The sign of the dephasing strength determines the variation of relative energy ${\rm sign}(\dot{\tilde{\varepsilon_t}})$ and entropy change. 
In particular, a positive dephasing strength yields  a monotonic increase of entropy. Indeed,   the rate of change  of the von Neumann entropy reads   
\begin{equation}
\frac{d S_t}{dt} = - \frac{\dot{u}_t}{(1-u_t)^2} \ln(u_t) = \gamma_t \frac{\varepsilon_t}{k_t^2}. 
\end{equation}
 Protocols restricted to $\gamma_t\geq 0$ allow only STA for thermalization to high-temperature states  (heating), with $\Delta S = S_f - S_0 >0$. Whenever values of  $\gamma_t$ can be engineered, this restriction is lifted. 

The maximum dephasing strength, illustrated in Fig.~\ref{fig3},  is specific to each scenario. It follows a different behavior when changing the trap frequency at constant temperature or vice versa. This is not surprising since the two parameters correspond to different physical phenomena, as discussed above.  The plateau observed when decreasing the temperature at a fixed trap frequency (Fig.~\ref{fig3}c) might be set by the trap size, which constrains the size of the particle. Interestingly, a given final phase-space density can be reached from different dephasing strengths, even when starting from a same initial state. In other words, processes yielding to $\beta_{f}\omega_{f}$ from $\beta_0 \omega_0$ can have different implementation costs according to $\gamma_{\rm{max}}$, as illustrated in Fig.~\ref{fig3}d. 

In conclusion, we have introduced shortcuts to adiabaticity {\aurelia with an open dynamics engineered to control the thermalization of a quantum oscillator.}  
The resulting  protocols are expected to be broadly applicable   as their implementation requires only a time modulation of the harmonic frequency and the dephasing strength,  accessible e.g. via stochastic parametric driving or continuous quantum measurements. 
Our results can be directly applied to non-Markovian dynamics whenever  the amplitude and sign of the dephasing strength can be engineered. 
{\aurelia Extension to obtain generalized Gibbs states and for interacting systems is under investigation. }

{\it Acknowledgments.---}  It is a pleasure to acknowledge  discussions with \'Angel Rivas and Bijay K. Agarwalla.


\appendix

\onecolumngrid

\section{Thermal state of a harmonic oscillator \label{appA}}
It is well known that the thermal state of a harmonic oscillator is Gaussian in the coordinate representation. For the sake of completeness, we briefly sketch the derivation below. 
For a time-independent harmonic oscillator, the Hamiltonian $\hat{H} = \frac{\hat{p}^2}{2m} + \frac{1}{2} m \omega^2 \hat{x}^2$ reads, in second quantization $\hbar \omega (a\dg a + \frac{1}{2})$, where $a= \sqrt{m \omega / 2\hbar} \hat{x} + i \sqrt{1/2m\hbar \omega} \hat{p}$ is the annihilation operator. 
The coordinate representation of the thermal operator $e^{- \beta \hat{H}}/Z$ is easily written using the Fock states, defined as $a \ket{n} = \sqrt{n} \ket{n-1}$, and reads  
\begin{eqnarray}
\rho(x,x') = \bra{x} \frac{e^{-\beta \hat{H}}}{Z} \ket{x'} = \frac{1}{Z } e^{- \beta \hbar \omega /2} \sum_n e^{-\beta \hbar \omega n} \braket{x}{n} \braket{n}{x'}.
\end{eqnarray}
Solving the Schr\"odinger equation for the Fock state wave function $\braket{x}{n}$ gives 
\begin{equation}
\braket{x}{n} = \sqrt{\frac{k}{ 2^n n! \sqrt{\pi}}} e^{-\frac{1}{2} (k x)^2} {H}_n\left(k x\right),
\end{equation}
where $k^{-1}= \sqrt{\hbar/ (m\omega)}$ denotes an effective length characteristic of the harmonic oscillator. 
The coordinate representation of the thermal density operator then reads
\begin{equation}
\rho(x,x') = \frac{k}{Z \sqrt{\pi} } e^{- \beta \hbar \omega /2}\exp\left({- \frac{k^2}{2}(x^2 + x'^2)}\right) \sum_n \left(\frac{e^{-\beta \hbar \omega}}{2}\right)^n \frac{1}{n!} {H}_n(k x) {H}_n(k x'). 
\end{equation}
We use Mehler's formula \cite{Mehler1866}, 
\begin{equation}  \label{eq:Mehler}
 \sum_{n=0}^\infty \frac{u^n}{2^n n!}{H}_n(x){H}_n(y)=\frac{1}{\sqrt{1-u^2}}e^{2 uxy/(1-u^2)}e^{-u^2(x^2+y^2)/(1-u^2)},
\end{equation}
 to  rewrite the sum with the Hermite polynomials as a Gaussian, yielding  
\begin{equation}\label{S5}
\rho(x,x') = \frac{k \sqrt{u}}{Z  \sqrt{\pi} }  \exp\left({-(k^2/2) (x^2 + x'^2)}\right)  \frac{1}{\sqrt{1-u^2}} \exp\left( - \frac{k^2}{(1-u^2)}(u^2 (x^2 + x'^2) - 2 u x x')\right), 
\end{equation}
where we have defined $u=e^{-\beta \hbar \omega}$. 
Finally, with the explicit form of the partition function $Z = {\rm Tr}(e^{- \beta \hat{H}})=1/(1-u)$, this coordinate representation also takes the form 
\begin{eqnarray}\label{S6}
\rho(x,x') = \frac{k}{ \sqrt{\pi}}\sqrt{\tanh (\beta \hbar \omega/2)}\exp\left( - (k^2/2) \coth (\beta \hbar \omega ) (x^2 + x'^2) + k^2 \sinh^{-1}(\beta \hbar \omega) x x'\right).
\end{eqnarray}
The same derivation holds for a time dependent Hamiltonian, and provides  the initial and final coefficients $A$ and $C$ given in the main text. We verify that the normalization factor is $ \frac{k}{ \sqrt{\pi}}\sqrt{\tanh (\beta \hbar \omega/2)}= \frac{k}{ \sqrt{\pi}}\sqrt{\frac{1-u}{1+u}}=\sqrt{2 (A+C)/\pi} $.

\section{Consistency equations from the evolution of the Gaussian Ansatz \label{appB}}
The master equation (\ref{eq:ME}) 
 in the coordinate representation reads
\begin{equation}\label{MEcoord}
\frac{d \rho_t(x,x')}{dt} =  \left( \frac{i \hbar}{2m}  \Big( \frac{\partial^2}{\partial x^2} - \frac{\partial^2}{\partial x'^2} \Big) - i \frac{m \omega_t^2}{2\hbar} (x^2 - x'^2) - \gamma_t (x - x')^2 \right) \rho_t(x,x').  
\end{equation}
For the Gaussian Ansatz given in Eq. (\ref{gaussansatz})
 of the main text, the real and imaginary parts of the evolution equation respectively give 
\begin{eqnarray}
    \frac{\dot{N}_t}{N_t} + \frac{2 \hbar}{m} B_t  &=& \left(\dot{A}_t +  \frac{4\hbar}{m}A_tB_t - \gamma_t\right)(x^2 + x'^2) +2 \left (\dot{C}_t + \frac{4 \hbar }{m} B_tC_t + \gamma_t\right) x x',\\
  \dot{B}_t&=&\frac{2\hbar}{m}(A_t^2-B_t^2-C_t^2)-\frac{m\omega_t^2}{2\hbar},
\end{eqnarray}
from which the consistency equations (\ref{consisteq})
 directly follow.

\section{Instantaneous diagonalization of the density matrix \label{appC}}
We look for the eigenvalues $p_{n,t}$ and eigenfunctions $\ket{\psi_{n,t}}$ that diagonalize the density matrix $\rho_t$ at any time. For the sake of simplicity, we omit the time dependence in the notation below. By definition, the eigenvalues fulfil $p_n >0$ and $\sum_n p_n = 1$, so we choose  to write them as $p_n = u^n (1-u)$, where $u$ can be seen as an exponential $e^{-\tilde{\varepsilon}}$.  We verify below that the functions  
\begin{equation}
 \braket{x}{\psi_n}  =\sqrt{\frac{{k}}{2^n n! \sqrt{\pi}}} e^{-\frac{{k}^2}{2} x^2} e^{ i B x^2}{H}_n ({k} x)
\end{equation}
correspond to the eigenfunctions. 
Note that orthogonality of the Hermite polynomials, $\int_\infty dx' {H}_n({k} x') {H}_m({k} x') e^{-{k} x'^2} =  \delta_{nm} 2^n n! \sqrt{\pi}/{k}$, guarantees orthonormality of the wave functions, $\braket{\psi_n}{\psi_m} = \delta_{nm}$. 
To justify the choice of this Ansatz and determine the time-dependent variables ${k}$ and $u$, we start with the coordinate representation 
\begin{equation}
\bra{x}\rho_t\ket{x'} = \sum_n u^n (1-u) \braket{x}{\psi_n} \braket{\psi_n}{x'},
\end{equation}
and use Mehler's equation  (\ref{eq:Mehler}) to get
\begin{equation}
\bra{x}\rho_t\ket{x'} = \frac{{k}}{\sqrt{\pi}}\sqrt{\frac{1-u}{1+u}} e^{- {k}^2 (x^2 + x'^2) (\frac{1}{2} + \frac{u^2}{1-u^2})} e^{i B (x^2 - x'^2)} e^{\frac{2 u {k}^2}{1-u^2}x x'}. 
\end{equation}
By identification, we obtain 
\begin{eqnarray}
A = {k}^2 \frac{1+u^2}{2 (1-u^2)}=\frac{{k}^2}{2} \coth \tilde{\varepsilon}, \qquad C = - {k}^2 \frac{u}{1-u^2}= -\frac{{k}^2}{2  \sinh\tilde{\varepsilon}},
\end{eqnarray}
the  reverse transformation corresponding to the physical setting being for $u>0$ and $0<-A/C<1$, and  
\begin{equation}
{k} = \Big(2 \sqrt{A^2 - C^2}\Big)^{1/2}, \qquad u = -\frac{A}{C} - \sqrt{\Big( \frac{A}{C}\Big)^2- 1}.
\end{equation}
The mean phonon number easily follows as $ \langle n\rangle = \sum_{n=0}^\infty n p_n=\frac{u}{1-u}$, and the  von Neumann entropy $S(\rho)=-\tr(\rho\log\rho)$ reads   
\begin{eqnarray}
  \label{eq:computentro}
 S(\rho)=-\sum_{n=0}^\infty p_n\ln p_n
  =-\frac{u\ln u}{1-u}-\ln(1-u)\,.
\end{eqnarray}

\section{Maximum dephasing strength \label{appD}}
We can show that the dephasing strength is inversely proportional to the time of the protocol for any polynomial Ansatz interpolating between the initial and final state. 
The time for which the dephasing strength is maximal is given from $\dot{\gamma}_{t_{\rm max}} = 0$, which leads
\begin{equation}
\left. \frac{d^2 f_\tau}{d^2 \tau} \right|_{t_{\rm max}}  = \left(\left.\frac{d f_\tau}{d\tau}\right|_{t_{\rm max}} \right)^2  (A_f - A_0 + C_f - C_0),
\end{equation}
where $\tau = t/t_f$ and where we have used $d f_\tau /dt = d f_\tau/ d\tau (1/t_f)$. 
This equation could be solved for a specific polynomial Ansatz $f_\tau = \sum_{n=0}^N f_n \tau^n$. Since $\gamma$ takes a zero value at initial and final time, a non-trivial solution goes through an extremum in the region $\tau \in [0:1]$. We denote the root corresponding to this time $r_1$. The maximum dephasing strength is reached at time $t_{max} = r_1 t_f$. We further have 
\begin{equation}
\gamma_{\rm max} = \frac{1}{t_f} \left. \frac{df_\tau}{d\tau} \right|_{t_{\rm max}} \frac{(A_f - A_0) C_{t_{\rm max}} - A_{t_{\rm max}} (C_f - C_0)}{A_{t_{\rm max}}+C_{t_{\rm max}}}.
\end{equation}
The parameters $A_{t_{\rm max}}$ and $C_{t_{\rm max}}$ are polynomials of $r_1$ and all terms on the r.h.s, apart from $1/t_f$,  depend only on the root $r_1$. This yields $\gamma_{\rm max} \propto 1/t_f$.

 \section{Determine the eigenfunctions overlap in Eq. (\ref{eq:srel})
  to evaluate the relative entropy \label{appE}}
We provide below the explicit form for the overlap 
\begin{eqnarray}
 &&\braket{m}{n}=\int_{-\infty}^{\infty}dx \braket{m}{x} \braket{x}{n}  = \frac{{k}}{ \sqrt{2^{n+m} n!m! \pi}} \int_{-\infty}^{\infty}dx e^{-{k}^2  x^2}   e^{- i B x^2}H_m\big({k} x\big) H_n ({k} x). \nonumber
 \end{eqnarray}
This overlap can be expressed as 
\begin{eqnarray}  \label{overlap}
 &&\braket{m}{n} = \frac{1}{ \sqrt{2^{n+m} n!m! \pi} }  I_{n,m}\left( 1 + i B / {k}^2 \right)  \end{eqnarray}
by defining the integral 
 \begin{equation}
 I_{n,m}(b) = \int_{-\infty}^{\infty}dx\, e^{-b x^2} {H}_n(x) {H}_m(x) , 
  \end{equation}
 where the indices $n$ and $m$ play a symmetric role. 
To solve this integral, we first write the exponential as $e^{-b x^2} = e^{- x^2}  e^{-(b-1) x^2} $ in order to have a Gaussian for each Hermite polynomial. Then, multiple integration by parts yield  $\int dx\, e^{-x^2} {H}_n(x) f(x) = \int dx\, e^{-x^2} D^n (g(x))$, where $D^n \equiv (d/ dx)^n$, for any function $g(x)$  \cite{delCampo2017a, Chenu2019b}. 
 So, for $g(x) = e^{-(b-1) x^2} {H}_m(x)$, and choosing $n<m$ by convention, we find 
 \begin{equation}
 I_{n,m}(b) = \int_{-\infty}^{\infty} dx\, e^{-x^2} D^n\big( e^{-(b-1) x^2} \big) {H}_m(x) .
 \end{equation}
We then expand the derivative $D^n$ in a binomial form, use the derivative of the Hermite polynomial, $D^j({H}_N(x) ) = 2^j N!/(N-j)! {H}_{N-j}(x)$, and the definition of the Hermite polynomial $e^{-x^2} {H}_N(x) = (-D)^N e^{-x^2}$ to obtain  
   \begin{eqnarray}
 I_{n,m}(b) &=& \sum_{l=0}^n \binom{n}{l} \frac{2^{n-l} m!}{(m-n+l)!}\int_{-\infty}^{\infty}dx e^{-x^2} D^l\big( e^{-(b-1) x^2} \big) {H}_{m-n+l}(x)\nonumber \\
 &=& \sum_{l=0}^n \binom{n}{l} \frac{2^{n-l} m! (-1)^{m-n+l} }{(m-n+l)!} \int_{-\infty}^{\infty}dx  D^l\big( e^{-(b-1) x^2} \big) D^{m-n+l}\big( e^{-x^2} \big) . 
 \end{eqnarray}
In order to evaluate the new integral, we write each of the derivates as a Fourier transform, using $D^n(e^{-x^2}) = (2i)^n / \sqrt{\pi} \int_\infty dt\, e^{-t^2} tn e^{2 i x t}$  \cite{AndrewsBook},
 which gives, taking $\alpha =( 1-b)$ and $j=(m-n+l)$,
    \begin{eqnarray}\label{eq:S27}
 \int_{-\infty}^{\infty}dx\,  D^l\big( e^{-\alpha x^2} \big) D^{j}\big( e^{-x^2} \big) &=&\frac{(2i)^{l+j}\sqrt{\alpha}^l}{\pi}  \int_{-\infty}^{\infty}dt ds \, e^{t^2 + s^2} t^l s^l\int_{-\infty}^{\infty}dx \, e^{2 i x (\sqrt{\alpha} t + s)} \nonumber\\
&=& (-1)^j  (2i \sqrt{\alpha})^{l+j}\int_{-\infty}^{\infty}dt \, e^{-t^2(1+\alpha)} t^{l+j}  \nonumber \\
&=&  (-1)^j  (2i \sqrt{\alpha})^{l+j}  \frac{1}{2} (1+\alpha)^{-\frac{l+j+1}{2}}(1+(-1)^{l+j}) \Gamma\big(\frac{l+j+1}{2}\big).
 \end{eqnarray}
 This leads to 
 \begin{equation}
I_{n,m}(b)  = (1 + (-1)^{m - n}) \sum_{l=0}^n  2^{m + l - 1} \binom{n}{l} \frac{m!}{(m - n + l)!} i^{m - n + 2 l} \left(1-\frac{1}{b}\right)^{\frac{2 l + m - n}{2}} \frac{1}{\sqrt{b}} \: \Gamma\left(\frac{1 + m - n}{2}+ l\right). 
\end{equation}
 We can further simplify this expression by noting that it is non-zero only for $(l+j) =( m-n+2l)$ even. Choosing $m>n$ by convention, we thus find $ I_{n,n+2p+1} =0$ for all integers $p$, and 
  \begin{equation} \label{I1}
 I_{n,n+2p}(b) = \sqrt{\frac{\pi}{b}}\sum_{l=0}^n \binom{n}{l}  2^{n-l}\left( \frac{1-b}{b}\right)^{p+l} \frac{(2p+n)! (2p+2l)! }{(2p+l)! (l+p)! }, 
  \end{equation}
  where we have used $\Gamma(1/2 + n) = \sqrt{\pi} (2n)!/(4^n n!) $ to explicitly write the Gamma function from Eq. (\ref{eq:S27}). 
Note that this sum can also be written using the hypergeometric function $_2F_1$, specifically 
    \begin{equation}
 I_{n,n+2p}(b) =2^n \sqrt{\frac{\pi}{b}}\left(\frac{1-b}{b} \right)^p  \frac{(2p+n)!}{p!} {}_2F_1\left(\frac{1}{2}+p,-n,1+2p,2-\frac{2}{b}\right). 
 \end{equation}
  Using this expression or Eq. (\ref{I1}) in (\ref{overlap}) with $b=1+ i B/{k^2}$ yields the overlap of interest.


\nocite{apsrev41Control}


%

\end{document}